\documentclass[seceq,preprint]{ptptex}

\usepackage{epsfig, color,graphicx}
\usepackage{amsmath}
\usepackage{amssymb}
\newcommand{\nn}{\nonumber \\}
\newcommand{\bea}{\begin{eqnarray}}
\newcommand{\ena}{\end{eqnarray}}

\newcommand{\hs}[1]{\hspace{#1 mm}}
\renewcommand{\a}{\alpha}

\renewcommand{\c}{\gamma}
\renewcommand{\d}{\delta}

\newcommand{\s}{\sigma}

\newcommand{\la}{\lambda}

\newcommand{\p}[1]{(\ref{#1})}

\newcommand{\tr}{\tilde{r}}

\newcommand{\squaret}{\kern1pt\vbox{\hrule height 0.9pt\hbox{\vrule width
0.9pt\hskip 2pt\vbox{\vskip 5.5pt}\hskip 3pt\vrule width 0.3pt}\hrule height
0.3pt}\kern1pt}
\begin{document}

%
%\preprint{KU-TP 04?}
\preprintnumber[3cm]{
KU-TP 044}
%
% title
%
\title{Global Structure of Black Holes in String Theory with Gauss-Bonnet Correction
in Various Dimensions
}
%%
%% author
%%
\author{
Nobuyoshi {\sc Ohta}$^{a,}$\footnote{e-mail address: ohtan at phys.kindai.ac.jp}
and Takashi {\sc Torii}$^{b,}$\footnote{e-mail address: torii at ge.oit.ac.jp}
}

\inst{
$^a$Department of Physics, Kinki University, Higashi-Osaka,
Osaka 577-8502, Japan\\
$^b$Department of General Education, Osaka Institute of Technology,
Asahi-ku, Osaka 535-8585, Japan
}
%%
%% abstract
%%
\abst{
We study global structures of black hole solutions in Einstein gravity with
Gauss-Bonnet term coupled to dilaton in various dimensions.
In particular we focus on the problem whether the singularity is weakened
due to the Gauss-Bonnet term and dilaton. We find that there appears the
non-central
singularity between horizon and the center in many cases, where the metric does not
diverge but the Kretschmann invariant does diverge. Hence this is a singularity,
but we find the singularity is much milder than the Schwarzschild solution
and the non-dilatonic one.
We discuss the origin of this ``fat'' singularity.
In other cases, we encounter singularity at the center which is much stronger
than the usual one.
We find that our black hole solutions have three different types of
the global structures;
the Schwarzschild, Schwarzschild-AdS and  ``regular AdS black hole" types.
}

\maketitle

\section{Introduction}

One of the long-standing problems in theoretical physics is how to reconcile
gravity with quantum theory. It is well known that Einstein gravity is not
renormalizable and there is an intrinsically difficult problem how to understand
physical properties near the singularity which always exists in black hole spacetimes
given as solutions in Einstein gravity. Superstring theory is the leading candidate
for quantum gravity. It is long expected that the theory could resolve
this problem. In order to study the geometrical properties and strong gravitational
phenomena, it is still difficult to apply full superstring theory itself.
In this situation,
it is appropriate to investigate these problems by using the effective low-energy
field theories including string quantum corrections.

Many works have been done on black hole solutions in dilatonic gravity,
and various properties have been studied since the work in Refs.~\citen{GM} and \citen{GHS}.
On the other hand, it is known that there are higher-order quantum corrections from
string theories.~\cite{MT} It is thus important to ask how these corrections may modify
the results. Several works have studied the effects of higher order
terms~\cite{KMRTW,AP,Alexeyev,TYM,CGO1,CGO2},
but most of the work considers theories without dilaton~\cite{BD,GG,Cai,TM,TM2},
which is one of the most important ingredients in the string effective theories.
Hence it is most significant to study black hole solutions and their properties
in the theory with the higher order corrections and dilaton.
The simplest higher order correction is the Gauss-Bonnet (GB) term
coupled to dilaton in heterotic string theories.

In our previous paper~\cite{GOT1}, we have studied asymptotically flat black
hole solutions with the GB correction term and dilaton without a cosmological constant
in various dimensions from 4 to 10 with $(D-2)$-dimensional hypersurface of
curvature signature $k=+1$.
We have then presented our results on black hole solutions with the cosmological
constant with $(D-2)$-dimensional hypersurface with $k=0, \pm 1$~\cite{GOT2,OT3,OT4}.
In the string perspective, it is also interesting to examine asymptotically
non-flat black hole solutions with possible application to AdS/CFT and dS/CFT
correspondences in mind~\cite{Mald,Stro,CNOS}.
Discussions of the origin of such cosmological constant are given in Refs.~\citen{AGMV}
and \citen{POL}.
Extremal solutions in similar systems have been discussed recently in Ref.~\citen{CGOO},
and other black hole solutions are presented in Ref.~\citen{MOS1}.% and \citen{MOS2}.

Most of the above works consider the properties of the external spacetime of the
black hole horizon except for Refs.~\citen{AP,Alexeyev,TM} and \citen{TM2}.
It is expected that the higher order corrections and dilaton significantly affects
the internal structure of spacetime and the structures of singularity.
For example, for the so-called small black holes, which have the singularity
at the horizon without enough number of charges to support the horizon,
it is known that the horizon is stretched due to the presence of
higher order terms, and the singularity is resolved there~\cite{Sen1,Sen2,Pres,CCMOP}.
The cosmological constant, on the other hand, is not expected to modify significantly
the short distance properties of the black holes, not affecting the singularities
much.
In this paper, we study this problem for the solutions we already obtained
in our series of papers~\cite{GOT1,GOT2,OT3,OT4}.
The black hole solutions exist for the cases
$(\Lambda,\; k)=(0,1),\; (1,1),\;(-1,0),\;(-1,1),\;(-1,1)$,
where $\Lambda$ is normalized to $0$ and $\pm 1$.
Since the external structures of  these black holes are studied already
in these papers, we focus on the internal properties in this paper.
We have found that cosmological horizons are not formed in the presence of the
positive cosmological constant~\cite{OT4}. Hence for the solutions with the cosmological
constant, we mainly discuss asymptotically anti-de Sitter (AdS) solutions with the
negative cosmological constant.

The most interesting problem is whether the higher order corrections and dilaton
make the strength of the singularity weaker. A special feature which occurs in the
presence of the GB term without dilaton is that the  singularity arises in the
intermediate position inside the horizon before reaching the center $r=0$ for the black
holes with negative mass~\cite{TM}. The singularity appears slightly weaker than
the Schwarzschild solutions. In other cases, the singularity appears at the origin,
and the singularity is weaker for $D\geq 5$ (the GB term does not give any effect
in $D=4$).
In our dilatonic case, we also find the similar kind of singularity for
lower-dimensional cases ($D=4,5$) in asymptotically flat solutions and
in $k=+1$ and 0 asymptotically AdS solutions.
We call this ``fat singularity'' because it is made fat due to the presence of the
GB term and dilaton. However, the formation mechanism of the fat singularity
is different from the non-dilatonic one.
It is deeply related to the singularity associated with dilaton.
In $D=5$ dimension, the fat singularity occurs for large black holes.
The metric itself does not diverge there while Kretschmann invariant diverges.
Hence this is a singularity and the spacetime ends there. The singularity itself
is not strong compared with the usual central singularity of Schwarzschild solutions
and the non-dilatonic case.
In all these cases, the singularities are spacelike. In other cases, we encounter
the singularity at the origin $r=0$, just like Schwarzschild solutions.
The nature of the central singularity deviates significantly from
the non-dilatonic case and the strength of the central singularity
is stronger than Schwarzschild solution.
We should note that these results are obtained for a particular value
of the dilaton coupling $\c=1/2$. Actually we confirm that the occurrence of the
fat singularity seems to depend on the dilaton coupling in the GB term.
If we adopt the other dilaton coupling $\c=1$, we find that the fat singularity
appears in $D=6$. Note also that the fat singularity always appears in four dimensions.
Thus the presence of the dilaton significantly affects the singularity.

This paper is organized as follows.
In \S~\ref{sec2}, we summarize the action and field equations of the theory we discuss.
In \S~\ref{IS}, we summarize the internal spacetime and global structure of
the black holes for the known cases. First in \S~\ref{GR}, we summarize the singularity
in Schwarzschild solution in general relativity (GR). Then we discuss the non-dilatonic
case with the GB term in \S~\ref{NDC}.
We go on to study the internal spacetime and the global structure of the dilatonic
solutions in \S~\ref{GS}. We give the detailed account of these for asymptotically
flat solutions in \S~\ref{L0k1}, for $\Lambda<0$ and $k=0$ case in \S~\ref{Lnk0},
for $\Lambda<0$ and $k=1$ in \S~\ref{Lnk1}, and finally for $\Lambda<0$ and $k=-1$
case in \S~\ref{Lnk-1}. In particular, we discuss the origin of the fat singularity.
Our conclusions and discussions are given in \S~\ref{CD}.

%%%%%%%%%%%%%%%%%%%%%%%%%%%%%%%%%%%%
%%%%%%%%%%%%%%%%%%%%%%%%%%%%%%%%%%%%
\section{Dilatonic Einstein-Gauss-Bonnet theory}
\label{sec2}
%%%%%%%%%%%%%%%%%%%%%%%%%%%%%%%%%%%%
%%%%%%%%%%%%%%%%%%%%%%%%%%%%%%%%%%%%

%%%%%%%%%%%%%%%%%%%%%%%%%%%%%%%%%%%%
%\subsection{Action and basic equations}
%%%%%%%%%%%%%%%%%%%%%%%%%%%%%%%%%%%%

We consider the following low-energy effective action for a
heterotic string
\bea %----------------
S=\frac{1}{2\kappa_D^2}\int d^Dx \sqrt{-g} \left[R - \frac12
 (\partial_\mu \phi)^2
 + \a_2 e^{-\c \phi} R^2_{\rm GB} -\Lambda e^{\lambda\phi}\right],
\label{act}
\ena %----------------
where $\kappa_D^2$ is a $D$-dimensional gravitational constant,
$\phi$ is a dilaton field, $\alpha_2=\a'/8$ is a numerical
coefficient given in terms of the Regge slope parameter $\alpha'$,
and
$R^2_{\rm GB} = R_{\mu\nu\rho\sigma} R^{\mu\nu\rho\sigma}
- 4 R_{\mu\nu} R^{\mu\nu} + R^2$ is the GB correction.
In this paper we take the coupling constant of dilaton $\c=1/2$,
the value that the ten-dimensional critical string theory predicts.

We parametrize the metric as
\bea
ds_D^2 = - B e^{-2\d} dt^2 + B^{-1} dr^2 + r^2 h_{ij}dx^i dx^j,
\ena
where $h_{ij}dx^i dx^j$ represents the line element of a
$(D-2)$-dimensional hypersurface with constant curvature
$(D-2)(D-3)k$ and volume $\Sigma_k$ for $k=\pm 1,0$.

The metric function $B=B(r)$ and the lapse function $\d=\d(r)$ depend only on the
coordinate $r$. The field equations are~\cite{GOT2,OT3}
\bea
&& \bigl[(k-B)\tr^{D-3}\bigr]' \frac{D-2}{\tr^{D-4}}h -\frac12 B \tr^2 {\phi'}^2
 - (D-1)_4\,e^{-\c\phi}\frac{(k-B)^2}{\tr^2} \nn
&& \hs{10} + 4(D-2)_3\, \c e^{-\c\phi}B(k-B)(\phi''-\c {\phi'}^2) \nn
&& \hs{20} + 2(D-2)_3\,\c e^{-\c\phi}\phi'\frac{(k-B)[(D-3)k-(D-1)B]}{\tr}
-\tr^2 \tilde{\Lambda}e^{\lambda\phi}
= 0\,,
\label{fe1} \\
&& \delta'(D-2)\tr h + \frac12 \tr^2 {\phi'}^2
 -2(D-2)_3\, \c e^{-\c\phi}(k-B)(\phi''-\c {\phi'}^2) =0 \,,
 \label{fe2} \\
&&
(e^{-\d} \tr^{D-2} B \phi')' = \c (D-2)_3 e^{-\c\phi-\d} \tr^{D-4}
\Big[ (D-4)_5 \frac{(k-B)^2}{\tr^2} + 2(B'-2\d' B)B' \nn
&& \hs{30} -4(k-B)BU(r) -4\frac{D-4}{\tr}(B'-\d'B)(k-B) \Big]
+e^{-\delta}\tr^{D-2}\lambda \tilde{\Lambda}e^{\lambda\phi},
\label{fe3}
\ena
where we have defined the dimensionless variables: $\tr \equiv r/\sqrt{\a_2}$,
$\tilde \Lambda = \a_2 \Lambda$, and the primes in the field equations
denote the derivatives with respect to $\tr$. Namely we measure our length
in the unit of $\sqrt{\a_2}$.
We will introduce the AdS radius $\ell$ and the mass
of the black hole $M_0$, which are renormalized as $\tilde{\ell}=\ell/\sqrt{\alpha'}$
and $\tilde{M}_0=M_0/{\alpha'}^{\frac{D-3}{2}}$, respectively.
In what follows, we omit tilde on the variables for simplicity.
We have also defined
\bea
(D-m)_n &\equiv& (D-m)(D-m-1)(D-m-2)\cdots(D-n), \nn
\label{h-def}
h &\equiv& 1+2(D-3) e^{-\c\phi} \Big[ (D-4) \frac{k-B}{r^2}
 + \c \phi'\frac{3B-k}{r}\Big], \\
\label{tilh-def}
\tilde h &\equiv& 1+2(D-3) e^{-\c\phi} \Big[(D-4)\frac{k-B}{r^2}
+\c\phi'\frac{2B}{r} \Big],
\ena
\bea
U(r) &\equiv& (2 \tilde h)^{-1} \Bigg[ (D-3)_4 \frac{k-B}{r^2 B}
-2\frac{D-3}{r}\Big(\frac{B'}{B}-\d'\Big) -\frac12 \phi'^2 \nn
&& + (D-3)e^{-\c\phi} \Bigg\{ (D-4)_6 \frac{(k-B)^2}{r^4 B}
- 4 (D-4)_5 \frac{k-B}{r^3}\Big(\frac{B'}{B}-\d'-\c\phi'\Big) \nn
&& -4(D-4)\c \frac{k-B}{r^2}\Big( \c \phi'^2 +\frac{D-2}{r}\phi'-\Phi \Big)
+8 \frac{\c\phi'}{r} \biggl[\Big(\frac{B'}{2}-\d' B\Big)\Big(\c\phi'-\d'
+\frac{2}{r} \Big) \nn
&& -\frac{D-4}{2r}B' \biggr] + 4(D-4)\Big(\frac{B'}{2B}-\d' \Big)
\frac{B'}{r^2}-\frac{4\c}{r}\Phi (B'-2\d'B)\Bigg\}
-\frac{1}{B} {\Lambda}e^{\lambda\phi}\Biggr],
\\
\Phi &\equiv& \phi'' +\Big(\frac{B'}{B}-\d' +\frac{D-2}{r}\Big) \phi'.
\label{dil}
\ena
This is the system we study in this paper.

%%%%%%%%%%%%%%%%%%%%%%%%%%%%%%%%%%%%%%%%%%%%%%%%%%%%%%%%%%%%%%%%
\section{Internal Spacetime and Global Structure in GR and Non-dilatonic case}
\label{IS}
%%%%%%%%%%%%%%%%%%%%%%%%%%%%%%%%%%%%%%%%%%%%%%%%%%%%%%%%%%%%%%%%

%%%%%%%%%%%%%%%%%%%%%%%%%%%%%%%%%%%%
\subsection{General Relativity}
\label{GR}
%%%%%%%%%%%%%%%%%%%%%%%%%%%%%%%%%%%%

In the limit of $\alpha\to 0$, Gauss-Bonnet gravity reduces to GR.
By the no-scalar hair theorem in GR,
the dilaton field becomes trivial. In this limit,
%===========<Equation>============%
\begin{eqnarray}
\label{einstein}
B=k-\frac{2M}{r^{D-3}}+\frac{r^2}{{\ell}^2},
\;\;\;\;\;
\delta \equiv 0,
\end{eqnarray}
%=================================%
is obtained. Here ${\Lambda}=-(D-1)_2/{\ell}^2$, and ${M}$ is
an integration constant related to the mass of the black hole.
In GR, the $(D-2)$-dimensional constant curvature space can be replaced by any
$(D-2)$-dimensional Einstein space.

The global structure of the spacetime is characterized by the properties of the
singularities, horizons, and infinity. Details of the global structure of the
solution in GR are discussed in Ref.~\citen{TM}.
There are six different types of global structure of the black hole solution.
They are Schwarzschild, Schwarzschild-dS, Schwarzschild-AdS, 
Reissner-Nordstr\"om-AdS, and extreme Reissner-Nordstr\"om-AdS types.
The remaining one is similar to the Schwarzschild-AdS type but 
the central singularity is replaced by the regular center. In this paper
we call it ``regular AdS black hole" type.

Here we summarize the behavior of the curvature around the singularity for comparison.
There is a curvature singularity at the center ($r=0$) except for the case with
${M}=0$.
Around the center, the Kretschmann invariant behaves as follows:
%===========<Equation>============%
\begin{eqnarray}
{\cal I}&=&R^{\mu\nu\rho\sigma}R_{\mu\nu\rho\sigma} \nn
&=& \bigl[B''-3B'\d'+2B(\d'^2-\d'')\bigr]^2 +\frac{2(D-2)}{r^2} (B'^2-2BB'\d'+2B^2 \d'^2)\nn
&& \hs{30} + \frac{2(D-2)_3}{r^4} (k-B)^2
\label{Kretch}
\\
&\sim&
O\Bigl(\frac{M^2}{r^{2D-2}}\Bigr).
\label{kretchemann_e}
\end{eqnarray}
%=================================%
When $M=0$, the black hole solution exists only for $k=-1$. The Kretschmann invariant
is finite at the center $r=0$ and the spacetime could be regular there.
%When the $(D-2)$-dimensional constant curvature space
%is simply connected, the center is regular. When it is not, as with the topological black
%holes, the center becomes singular. This singularity is simply due to the
%topological structure of the constant curvature space.

%%%%%%%%%%%%%%%%%%%%%%%%%%%%%%%%%%%%
\subsection{Non-dilatonic Case}
\label{NDC}
%%%%%%%%%%%%%%%%%%%%%%%%%%%%%%%%%%%%

In the non-dilatonic case ($\gamma=0$ and $\phi\equiv 0$), the GB term becomes
topological invariant and does not contribute to the field equation for $D=4$. Hence
we consider the $D\geq 5$ cases. The gravitational equations give the solution
%===========<Equation>============%
\begin{equation}
\label{f-eq}
B(r) =k+ \frac{1}{2(D-3)_4}
 \left(1 \mp \sqrt{1-\frac{4(D-3)_4}{{\ell}^2} + \frac{8(D-3)_4\bar{M}}{r^{D-1}}}\,
\right) r^{2},
\;\;\;\;\;
\delta \equiv 0,
\end{equation}
%=================================%
where ${\Lambda}=-(D-1)_2/{\ell}^2$, and  $\bar{M}$ is an integration constant
related to the mass of the black hole.
This is the extended Boulware-Deser solution including the cosmological constant
and  the topological black hole case~\cite{Cai}.
There are two families of solutions depending on the sign in front of the square root in
Eq.~(\ref{f-eq}). We call the solution with minus (plus) sign the GR (GB) branch solution
because the solution with minus sign reproduces the solution in GR in the $\alpha \to 0$ limit
while there is no such limit for the solution with plus sign.
Details of the global structure of this solution are summarized in Ref.~\citen{TM}.
Just as in the GR case, there are six different types of global structure of
the black hole solutions.

Besides the central singularity at $r=0$, there can be another type of singularity
at $r=r_b>0$ called the branch singularity. The value $r_b$ is obtained by the condition
that the square root term in Eq.~(\ref{f-eq}) vanishes, i.e., two branches of
solutions coincide.
In order for the solution to be well defined, the following condition should be satisfied:
%===========<Equation>============%
\begin{eqnarray}
\label{vacuum}
\frac{4(D-3)_4}{{\ell}^2} < 1.
\end{eqnarray}
%=================================%
Under this condition, there is no branch singularity when $\bar{M}\geq 0$,
while there exists the branch singularity for $\bar{M}< 0$. We find that
$r_b$ is given by
%===========<Equation>============%
\bea
r_b^{D-1}=8(D-3)_4 \biggl[1-\frac{4(D-3)_4}{ \ell^2}\biggr]^{-1} |\bar{M}| .
\ena
%=================================%

On the other hand, the positive-mass solutions have only a central singularity.
The metric function behaves as
%===========<Equation>============%
\begin{equation}
%\label{divGB}
B \approx  \mp \sqrt{\frac{2\bar{M}}{(D-3)_4r^{D-5}}}
+k+\frac{r^2}{2(D-3)_4}\mp \frac{1}{8}\sqrt{\frac{1}{2[(D-3)_4]^3\bar{M}}}
\biggl(1-\frac{4(D-3)_4}{\ell^2}\biggr)
r^{\frac{D+3}{2}},
\end{equation}
%=================================%
around the center.
The Kretschmann invariant behaves as
%===========<Equation>============%
\begin{eqnarray}
\label{divGB}
{\cal I}\sim
O\Bigl(\frac{\bar{M}}{r^{D-1}}\Bigr).
\end{eqnarray}
%=================================%
Although $B$ looks finite for $D=5$ and there appears to be no singularity at the center,
the Kretschmann invariant diverges as
%===========<Equation>============%
\begin{eqnarray}
%\label{??}
{\cal I}
\sim
\frac{6\bar{M}}{r^4}+O(r^{-2}),
\end{eqnarray}
%=================================%
which is the same order as Eq.~(\ref{divGB}), so that the center is singular also
in this case.

When the branch singularity exists, the metric function $B$ behaves around it as
%===========<Equation>============%
\begin{equation}
B(r) \approx \biggl(k+\frac{r_b^2}{2(D-3)_4}\biggl)
\mp
\frac{r_b^2}{2 (D-3)_4}\sqrt{\frac{D-1}{r_b}\biggl(1-\frac{4(D-3)_4}{\ell^2}\biggl)}
(r-r_b)^{1/2},
\end{equation}
%=================================%
and the Kretschmann invariant
behaves as
%===========<Equation>============%
\begin{equation}
\label{divbranch}
{\cal I}\sim O\biggl[\frac{1}{(r-r_b)^{3}}\biggr].
\end{equation}
%=================================%
It is interesting to note that the divergent rate does not depend on the dimensions,
in contrast to the central singularity.

Note that the divergent behavior of the central singularity in Gauss-Bonnet gravity
is milder than that in GR [compare Eqs.~(\ref{kretchemann_e}) and
(\ref{divGB})]. We also point out that the divergent behavior
of the branch singularity is milder than that of the central
singularity [compare Eqs.~(\ref{divbranch}) and (\ref{divGB})].
The global structures of the non-dilatonic black holes are
classified into six different types which are same as the 
GR case.

%%%%%%%%%%%%%%%%%%%%%%%%%%%%%%%%%%%%%%%%%%%%%%%%%%%%%%%%%%%%%%%%
\section{Internal Spacetime and Global Structure in Dilatonic GB Theory}
\label{GS}
%%%%%%%%%%%%%%%%%%%%%%%%%%%%%%%%%%%%%%%%%%%%%%%%%%%%%%%%%%%%%%%%

%%%%%%%%%%%%%%%%%%%%%%%%%%%%%%%%%%%%
\subsection{$\Lambda=0$ and $k=1$}
\label{L0k1}
%%%%%%%%%%%%%%%%%%%%%%%%%%%%%%%%%%%%

Let us proceed to the dilatonic black holes in GB gravity.
First one is the solutions with $k=1$ and no cosmological constant~\cite{GOT1}.
All of these solutions are asymptotically locally flat, and the mass of the black
hole $M_0$ is defined by
%===========<Equation>============%
\begin{eqnarray}
B \to k-\frac{2M_0}{r^{D-3}}~~~~ ({\rm as} ~r\to \infty).
\end{eqnarray}
%=================================%
There is no horizon
outside of the black hole event horizon. Hence the domain of outer communication
of the solution has the same structure as that of Schwarzschild black hole.
We will discuss  the internal structure of the solution individually for each dimension.

%===========================================
\subsubsection{$D=4$}
%===========================================

The internal structure of the dilatonic solution with $\gamma =1$ in $D=4$ was
investigated in Ref.~\citen{Alexeyev}. Here we examine the  $\gamma =1/2$ case.

Integrating the field equations inward from the black hole event horizon with
the boundary values which are obtained from the exterior solutions, we find that
for any size of black hole, the singularity exists at the nonzero finite radius
$r_{\rm s}$. The inside of this radius is disconnected from our world.
Around this singularity, the field functions behaves as
%===========<Equation>============%
\begin{eqnarray}
\hs{-5}
B-B(r_{\rm s}) \sim (r-r_{\rm s})^{1/2}, \;\;
\delta-\delta(r_{\rm s}) \sim  (r-r_{\rm s})^{1/2},\;\;
\phi-\phi(r_{\rm s})\sim (r-r_{\rm s})+  (r-r_{\rm s})^{3/2}.~~
\label{fields_r_s}
\end{eqnarray}
%=================================%
It should be noted that the metric functions and the dilaton field are finite even
at $r=r_{\rm s}$. These behaviors of the field functions around $r_{\rm s}$ are
universal for all the solution in all dimension as we will see later.
We call this singularity the fat singularity.
The Kretschmann invariant diverges as
%===========<Equation>============%
\begin{eqnarray}
%\label{??}
{\cal I}\sim O\biggl[\frac{1}{(r-r_{\rm s})^{4}}\biggr].
\end{eqnarray}
%=================================%
The leading behavior is governed by $k$ in the third term in Eq.~(\ref{Kretch}).
The divergence is slightly stronger than in the non-dilatonic case.
The mechanism of the appearance of the singularity at finite radius will
be discussed in the next $D=5$ case.

\begin{figure}[bt]
\begin{center}
\includegraphics[width=16cm]{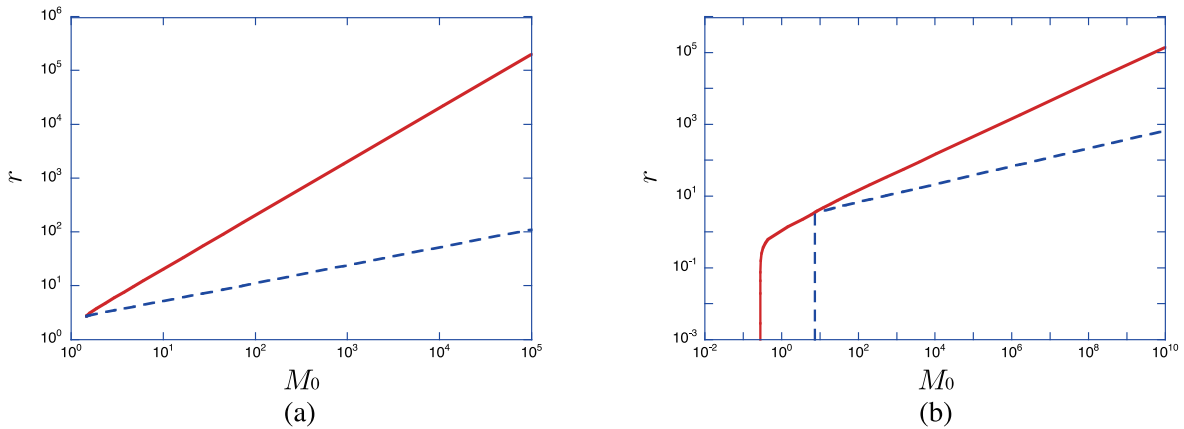}
\end{center}
\caption{The relations of $r_{\rm s}$ (dashed and blue line) and $r_H$ (solid and red line)
with ${M}_0$ for the black hole solutions in (a) 4 and (b) 5 dimensions with
$\gamma=1/2$, $\Lambda=0$, $k=1$.}
\label{M-rsing_L0k1D4}
\end{figure}

We show the relations of $r_s$ and $r_H$ with the black hole mass in
Fig.~\ref{M-rsing_L0k1D4}.
There is a lower limit on the size of the black hole in $D=4$. The second derivative
of the dilaton field diverges at the horizon for the minimum solution. Our analysis
shows that the horizon radius and radius of the fat singularity coincide
at this limiting size. This means that the existence of the lower limit is due to
the appearance of the  singularity. Further analysis is necessary to determine
whether the singularity of the minimum solution is naked or not.
The locations of the $r_H$ and $r_{\rm s}$ depend on the black hole mass ${M}_0$ as
$r_H \propto {M}_0$ and $r_{\rm s} \propto {M}_0^{1/3}$, respectively
(See Fig.~\ref{M-rsing_L0k1D4}~(a)).
The physical interpretation of this dependence is under investigation.

%--figures-------------------------------------------------------------
%--figures-------------------------------------------------------------

The global structure of the spacetime is determined by the properties of the
singularity, horizons and infinities. Whether the singularity is spacelike, null,
or timelike depends on the dominant term of the metric function $B$ near the
singularity. The tortoise coordinate is defined by
%===========<Equation>============%
\begin{eqnarray}
%\label{??}
r_{\ast}=\int^{r} \! e^{\delta} B^{-1} dr.
\end{eqnarray}
%=================================%
Since the metric function $B$ is negative and the tortoise coordinate is finite
at the singularity $r=r_s$, the singularity is spacelike. The spacetime
is asymptotically flat. There is no root of $B=0$ inside of the black hole horizon,
so no inner horizon. Hence the global structure of these solutions is the same
as that of Schwarzschild black hole.

%===========================================
\subsubsection{$D=5$}
%===========================================

%Although the solution has singularity for nonzero finite radius in $D=4$,
%it has none for $D=5$.
For the small black holes with $r_H<3.46$, the singularity exists
at the center and the spacetime is regular for $r>0$ as in the GR case.
Assuming that the power dependence of the functions at the center is given as
%===========<Equation>============%
\begin{eqnarray}
\label{function_div}
B\sim r^{a},
~~~
e^{2\delta} \sim r^{b},
~~~
e^{\gamma\phi}\sim r^{c},
\end{eqnarray}
%=================================%
and those terms including $B$ and $e^{\gamma \phi}$ in the field
equations~(\ref{fe1})--(\ref{fe3}) give the leading contributions,
we find that $a-c=2$ in arbitrary dimensions.
Our numerical analyses in $D=5$ indeed show that
the field functions behave as
%===========<Equation>============%
\begin{eqnarray}
\label{d5div}
B\sim r^{-10.9},
~~~
e^{2\delta} \sim r^{-12.9},
~~~
e^{\gamma\phi}\sim r^{-12.9},
\end{eqnarray}
%=================================%
which is independent of the horizon radius.
This means that the Kretschmann invariant diverges as
%===========<Equation>============%
\begin{eqnarray}
\label{d5violent}
{\cal I} \sim O\biggl(\frac{1}{r^{25.8}}\biggr).
\end{eqnarray}
%=================================%
This is violent divergence, and is much stronger than Schwarzschild solution
in GR.

Since the metric function $B$ is negative and the tortoise coordinate is finite
at the center, the central singularity is spacelike. There is no inner horizon.
Hence the global structure of these solutions is again the same as that of
Schwarzschild black hole.

\begin{figure}[bt]
\begin{center}
\includegraphics[width=16cm]{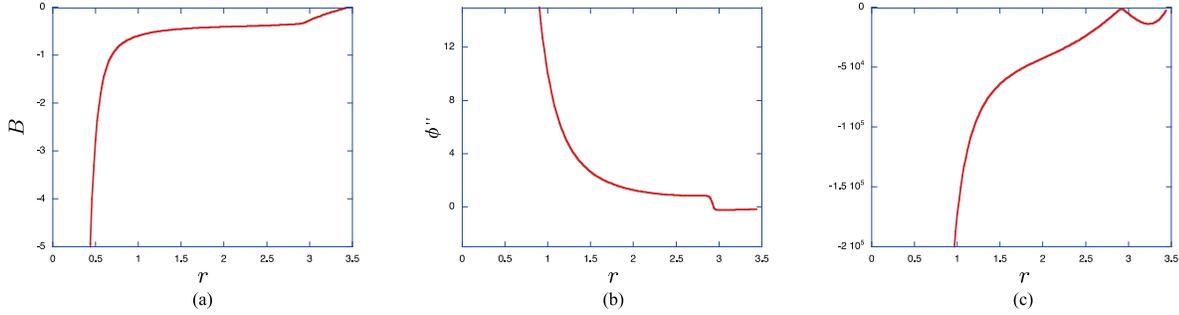}
\end{center}
\caption{The configurations of (a) the metric function $B$, (b) the second derivative
of the dilaton field
and (c) the denominator of the equation $\phi''$ for the solution in 5 dimensions
with $\gamma=1/2$, $\Lambda=0$, $k=1$, and $r_H=3.45$.
}
\label{d5k1_configDDphi}
\end{figure}

The behaviors of the field functions are shown in Fig.~\ref{d5k1_configDDphi}.
The metric function $B$ monotonically increases inside of the horizon.
When the horizon radius becomes as large as $r_H=3.46$,
the denominator in the equation of the dilaton field $\phi''= \cdots$ vanishes
at $r \approx 2.9$. In Fig.~\ref{d5k1_configDDphi}~(c), we show the behavior
of the denominator for the value of $r_H$ just before the singularity is formed.
At $r_H=3.46$, the second derivative of the dilaton field diverges and the
singularity is formed at this point. This is the fat singularity.
This fat singularity locates at nonzero finite radius
as the branch singularity in the non-dilatonic case. Its formation is, however,
caused by the different mechanism. It is due to the singularity associated with
the dilaton. We also see that there is no other solution,
hence this is not the branch singularity.

%--figures-------------------------------------------------------------
%--figures-------------------------------------------------------------

For the large black holes with $r_H>3.46$,
the singularity exists at the nonzero finite radius $r_{\rm s}$ as in the $D=4$ case.
Around the fat singularity, the field functions behaves as Eq.~(\ref{fields_r_s}).
The Kretschmann invariant diverges as
%===========<Equation>============%
\begin{eqnarray}
%\label{??}
{\cal I} \sim O\biggl[\frac{1}{(r-r_{\rm s})^{4}}\biggr].
\end{eqnarray}
%=================================%
This is the same divergent rate as the central singularity with positive mass
in the non-dilatonic case.
The locations of the $r_H$ and $r_{\rm s}$ depend on the black hole mass ${M}$ as
$r_H \propto {M}^{1/2}$ and $r_{\rm s} \propto {M}^{1/4}$, respectively
(See Fig.~\ref{M-rsing_L0k1D4}~(b)).
Since  the tortoise coordinate is finite
at $r=r_{\rm s}$, the singularity is spacelike. Hence the global structure of these solutions
is again the same as that of Schwarzschild black hole.

%===========================================
\subsubsection{$D=6$ -- $10$}
%===========================================
%--figures-------------------------------------------------------------
\begin{figure}[bt]
\begin{center}
\includegraphics[width=16cm]{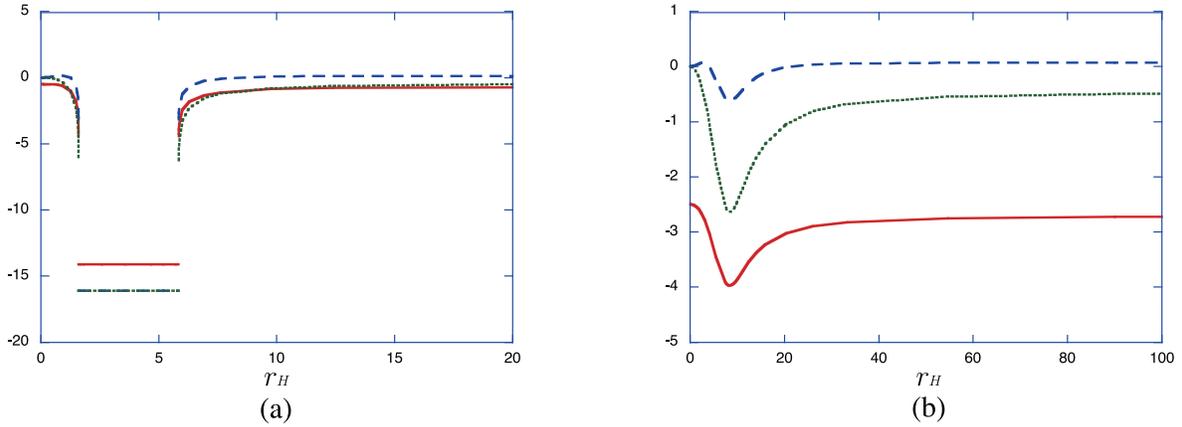}
\end{center}
\caption{The divergent rates of the field functions around the central singularity
for the black hole solution
in (a) 6 and (b) 10 dimensions with $\gamma=1/2$, $\Lambda=0$, $k=1$.
We plot the exponents of $B\sim r^{a}$ (solid and red line), $e^{2\delta}\sim r^{b}$
(dashed and blue line), and $e^{\gamma \phi}\sim r^{c}$ (dotted and green line).}
\label{rh-div_L0k1D610}
\end{figure}
%--figures-------------------------------------------------------------

In dimensions higher than five, we do not find the fat singularity for any
horizon radius.
The singularity exists at the center and the spacetime is regular for $r>0$.
There is no inner horizon. Assuming the behavior of the field functions as
Eq.~(\ref{function_div}), we plot the exponents $a,\:b,\:c$ as functions of $r_H$
in Fig.~\ref{rh-div_L0k1D610}.
The divergent rates depend on the horizon radius.
In $D=6$, there is the range $r_H\in (1.613,~5.848)$ where the exponents become constant as
%===========<Equation>============%
\begin{eqnarray}
\label{L0k1D6div}
B\sim r^{-14.1},
~~~
e^{2\delta} \sim r^{-16.1},
~~~
e^{\gamma\phi}\sim r^{-16.1}.
\end{eqnarray}
%=================================%
%For the black hole solution with $r_H\in (1.613,~5.848)$, the value of the dilaton field at the horizon is in the range $\phi_H\in (0.774,~-0.180)$.
We find the similar behavior in $D=7$:
%===========<Equation>============%
\begin{eqnarray}
\label{L0k1D7div}
B\sim r^{-17.3},
~~~
e^{2\delta} \sim r^{-19.3},
~~~
e^{\gamma \phi}\sim r^{-19.3}.
\end{eqnarray}
%=================================%
for the range $r_H\in (3.59,~5.77)$.
In this range, the exponents have the relation $a-2=c$
as in the $D=5$ case. By numerical analysis, we find that
%===========<Equation>============%
\begin{eqnarray}
\label{L0k1div_a}
a=-3.1807D+4.9683,
\end{eqnarray}
%=================================%
for these ranges.
For larger dimensions $D\geq 8$, however, there is no such range with constant exponents
as shown in Fig.~\ref{rh-div_L0k1D610}~(b).

Except for the ranges discussed above, the divergent rates depend on the horizon radius.
For the large black hole, the exponents approach $a=-0.67,~b=0.15,~c=-0.41$ ($D=6$),
$a=-2.72,~b=0.067,~c=-0.46$ ($D=10$).
For the function $B$, the mildest divergence appears in the zero horizon limit, where
$a=-(D-5)/2,~b=0,~c=0$.
The exponents changes discontinuously at $r = 1.613$ and  $r = 5.848$.

The Kretschmann invariant diverges as
%===========<Equation>============%
\begin{eqnarray}
%\label{??}
{\cal I} \sim O(r^{2a-4}).
\end{eqnarray}
%=================================%
Hence the Kretschmann invariant diverges most violently with the rate $r^{-32.2}$
in the range $r_H\in (2.686,~6.884)$ in $D=6$, and $r^{-11.9}$ for $r_H=8.26$ in $D=10$.
The mildest divergence is obtained in the zero horizon limit in both dimensions as
%===========<Equation>============%
\begin{eqnarray}
%\label{??}
{\cal I} \sim O\biggl(\frac{1}{r^{D-1}}\biggr),
\end{eqnarray}
%=================================%
which is the same order as the non-dilatonic one.

Since  the tortoise coordinate is finite
at $r=0$, the singularity is spacelike. There is no event horizon except for the black hole
horizon. The spacetime is asymptotically flat.
Hence the global structure of these solutions is again the Schwarzschild black hole type.

In this paper we consider the model with $\gamma =1/2$. However, depending on how
compactification is made and/or other circumstances, $\gamma$ can take different
values such as $\gamma =\sqrt{2/(D-2)}$.
We have confirmed that whether and when the fat singularity appears depends
on $\gamma$. For example, the fat singularity appears at $r=7.95307$ for
$D=6$, $\gamma=1$, and $r_H=8.49422$.
This suggests that the fat singularity is more likely to appear for the larger dilaton
coupling.

%%%%%%%%%%%%%%%%%%%%%%%%%%%%%%%%%%%%
%\subsection{$\Lambda=0$ and $k=0$}
%%%%%%%%%%%%%%%%%%%%%%%%%%%%%%%%%%%%
%No black hole solution!

%%%%%%%%%%%%%%%%%%%%%%%%%%%%%%%%%%%%
%\subsection{$\Lambda=0$ and $k=-1$}
%%%%%%%%%%%%%%%%%%%%%%%%%%%%%%%%%%%%
%No black hole solution!

%%%%%%%%%%%%%%%%%%%%%%%%%%%%%%%%%%%%
%\subsection{$\Lambda=1$ and $k=1$}
%%%%%%%%%%%%%%%%%%%%%%%%%%%%%%%%%%%%
%No black hole solution!

%%%%%%%%%%%%%%%%%%%%%%%%%%%%%%%%%%%%
%\subsection{$\Lambda=1$ and $k=0$}
%%%%%%%%%%%%%%%%%%%%%%%%%%%%%%%%%%%%
%No black hole solution!

%%%%%%%%%%%%%%%%%%%%%%%%%%%%%%%%%%%%
%\subsection{$\Lambda>0$ and $k=-1$ ~~ ($D\geq 5$) ~~{\color{blue}  [paper IV]}}
%%%%%%%%%%%%%%%%%%%%%%%%%%%%%%%%%%%%

%%%%%%%%%%%%%%%%%%%%%%%%%%%%%%%%%%%%
\subsection{$\Lambda <0 $ and $k=0$}
\label{Lnk0}
%%%%%%%%%%%%%%%%%%%%%%%%%%%%%%%%%%%%

For the solutions with negative cosmological constant~\cite{GOT2,OT3}, we impose the
``AdS asymptotic behavior" at infinity. It is
%===========<Equation>============%
\bea
\label{as}
B \sim {b}_2 r^2 +k - \frac{2 M}{r^{\mu}}, ~~~
\d(r) \sim \d_0 + \frac{\d_1}{r^{\s}}, ~~
\phi \sim \phi_0 + \frac{\phi_1}{r^{\nu}} \,,
\ena
%=================================%
with finite constants ${b}_2>0$, $ M$, $\d_0$, $\d_1$,  $\phi_0$, $\phi_1$
and positive constant $\mu$, $\s$, $\nu$.
The coefficient of the first term ${b}_2$ is related to the AdS radius
as ${b}_2 = \ell^{-2}$.
By analyzing the asymptotic expansion  \cite{GOT2,OT3}, we find
%===========<Equation>============%
\begin{eqnarray}
\label{AdS_B2}
{{b}_2}^{\;2}=\frac{\lambda|{\Lambda}|}{(D)_3 \gamma}
\biggl[\frac{D(D-3)}{(D-1)_2}\frac{\gamma|{\Lambda}|}{\lambda}
\biggl(1+\frac{(D-4)\lambda}{D\gamma}\biggr)^2
\biggr]^{\frac{\gamma+\lambda}{\gamma-\lambda}}.
\end{eqnarray}
%=================================%
Eq.~(\ref{as}) is not, however,  sufficient
for the spacetime to be the exactly AdS asymptotically. Strictly
speaking, the asymptotically AdS spacetime is left invariant under
$SO(D-1,2)$~\cite{Henneaux}. Whether the solution satisfies the AdS-invariant
boundary condition or not depends on the value of the power indices
$\mu$, $\s$, and $\nu$. The mass of the solution ${M}_0$ is defined
by\footnote{See Ref.~\citen{GOT2} for the details.}
%===========<Equation>============%
\begin{eqnarray}
%\label{??}
-g^{tt} \to \;{b}_2 r^2 +k-\frac{2{M}_0}{r^{D-3}}
~~~~({\rm as} ~ r\to \infty).
\end{eqnarray}
%=================================%

To study black hole solutions for various cosmological constants,
it is convenient to note that the field equations have a shift symmetry~\cite{GOT2,OT3}:
\bea
\phi \to \phi-\phi_{\ast}, ~~
\Lambda \to e^{(\la-\c)\phi_{\ast}} \Lambda, ~~
B \to e^{-\c \phi_{\ast}} B\, ,
\label{sym2}
\ena
where $\phi_{\ast}$ is an arbitrary constant.
This changes the magnitude of the cosmological constant. Hence this may
be used to generate solutions for different cosmological constants but
with the same horizon radius, given a solution for some cosmological constant
and $r_H$.
Besides the above symmetry, the field equations for $k=0$ are invariant under
the following scaling transformation:
\bea
B \to a^2 B, ~~
r \to a r,
\label{sym1}
\ena
with an arbitrary constant $a$.
If a black hole solution with the horizon radius $r_H$ is obtained,
we can generate solutions with different horizon radii but the same $\Lambda$
by this scaling transformation.
Combining these two symmetries, we can find relation between the mass and the horizon radius:
%===========<Equation>============%
\begin{eqnarray}
\label{M-rh_k=0}
{M}_0 \propto |\Lambda |^{\gamma/(\gamma-\lambda)} r_H^{D-1}.
\end{eqnarray}
%=================================%
These scaling symmetries also means
%===========<Equation>============%
\begin{eqnarray}
\label{rs-M_k=0}
{M}_0 \propto |\Lambda |^{\gamma/(\gamma-\lambda)} r_{\rm s}^{D-1},
%~~~~~
%\Bigl[\mbox{or  }
%r_s\propto \bigl( |\Lambda |^{(\gamma-\lambda)/\gamma} {M}_0 \bigr)^{1/(D-1)}\Bigr],
\end{eqnarray}
%=================================%
if there is a fat singularity at $r_{\rm s}$.
As in Ref.~\citen{GOT2} we fix the values of the parameters as $\gamma=1/2$, $\lambda=1/3$
in the following numerical analysis in this subsection.

%===========================================
\subsubsection{$D=4$ and $D=5$}
%===========================================

The exterior solution is obtained by integrating the field equations from the event
horizon with the suitable boundary condition to infinity.
Then we find the ${M}_0$-$r_H$ relation as
%===========<Equation>============%
\begin{eqnarray}
&&{M}_0 =0.0830\: |\Lambda |^{3} r_H^{3} ~~~ (D=4), \\
&&{M}_0 =0.140\: |\Lambda |^{3} r_H^{4} ~~~ (D=5).
\end{eqnarray}
%=================================%

On the other hand, by integrating inward from the event horizon, we find the fat
singularity at finite radius $r_s$ for any horizon radius.
Locations of the singularity are found to be
%===========<Equation>============%
\begin{eqnarray}
%\label{??}
&& r_{\rm s} =0.883 \:r_H ~~~ (D=4), \\
&& r_{\rm s} =0.827 \:r_H ~~~ (D=5).
\end{eqnarray}
%=================================%
%By Eq~(\ref{rs-M_k=0}), the dependence of $r_s$ on black hole mass is
%===========<Equation>============%
%\begin{eqnarray}
%\label{??}
%&& r_{\rm s} = ??????? \:\bigl(|\Lambda |^{1/3} {M}\bigr)^{1/3}  ~~~ (D=4), \\
%&& r_{\rm s} = ??????? \:\bigl(|\Lambda |^{1/3} {M}\bigr)^{1/4} ~~~ (D=5).
%\end{eqnarray}
%=================================%
The behaviors of the field functions around the fat singularity are the same as the
case with $\Lambda =0$ and $k=1$ given in Eq.~(\ref{fields_r_s}).
The Kretschmann invariant diverges as
%===========<Equation>============%
\begin{eqnarray}
%\label{??}
{\cal I} \sim O\biggl[\frac{1}{(r-r_{\rm s})^{3}}\biggr].
\end{eqnarray}
%=================================%
This divergence is milder than the non-dilatonic case. Since the singularity
locates at the finite tortoise coordinate, it is spacelike.
Then, the global structure is Schwarzschild-AdS type.

%===========================================
\subsubsection{$D=6-10$}
%===========================================

The ${M}_0$-$r_H$ relation is given by
%===========<Equation>============%
\begin{eqnarray}
&&{M}_0 =0.159\: |\Lambda |^{3} r_H^{5} ~~~ (D=6), \\
&&{M}_0 =0.0794\: |\Lambda |^{3} r_H^{9} ~~~ (D=10).
\end{eqnarray}
%=================================%
Integrating inward, we find that the singularity exists at the center, and the spacetime
is regular for $r>0$.
There is no inner horizon. The field functions behave as
%===========<Equation>============%
\begin{eqnarray}
\label{L-1k0D6_div}
&& B\sim r^{-14.1}, ~~~ e^{2\delta} \sim r^{-16.1}, ~~~ e^{\gamma\phi}\sim r^{-16.1}~~~(D=6),
 \\
&& B\sim r^{-26.8}, ~~~ e^{2\delta} \sim r^{-28.8},~~~e^{\gamma\phi}\sim r^{-28.8}~~~ (D=10).
\end{eqnarray}
%=================================%
These violent divergences are similar to the one expressed by Eqs.~(\ref{L0k1D6div})
and (\ref{L0k1D7div}) in the $\Lambda=0$ and $k=1$ case.
The exponent of $B$ is obtained and it is given by Eq.~(\ref{L0k1div_a}).
The Kretschmann invariant diverges as
%===========<Equation>============%
\begin{eqnarray}
%\label{??}
&& {\cal I} \sim O\biggl(\frac{1}{r^{32.2}}\biggr)~~~ (D=6),
\\
&& {\cal I} \sim O\biggl(\frac{1}{r^{57.6}}\biggr)~~~ (D=10).
\end{eqnarray}
%=================================%
The singular behavior at the center is quite strong.
The singularity is spacelike, and
the global structure is Schwarzschild-AdS type.

%%%%%%%%%%%%%%%%%%%%%%%%%%%%%%%%%%%%
\subsection{$\Lambda<0$ and $k=1$}
\label{Lnk1}
%%%%%%%%%%%%%%%%%%%%%%%%%%%%%%%%%%%%

As we will see below, the internal structures of the black hole solution with
$\Lambda<0$ and $k=1$~\cite{OT3} are qualitatively the same as those in the
$\Lambda=0$ and $k=1$ case. This implies that the cosmological constant
does not affect the internal structure so much as we expected.
All the black hole solutions are locally AdS at infinity.
The inverse square of the AdS radius ${b}_2$ is given by Eq.~(\ref{AdS_B2}).

%===========================================
\subsubsection{$D=4$}
%===========================================

\begin{figure}[bt]
\includegraphics[width=16cm]{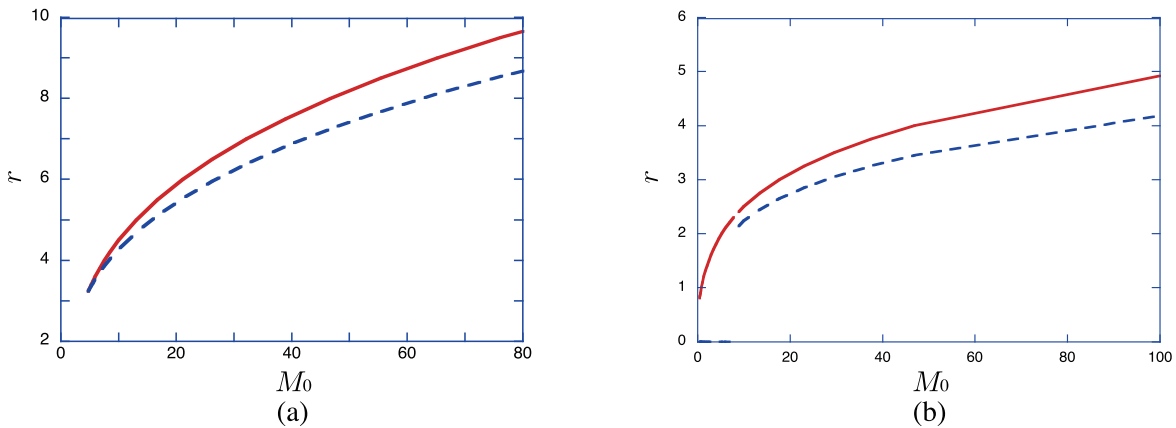}
\caption{The $r_{\rm s}$-${M}_0$ (dashed and blue line) and $r_H$-${M}_0$ (solid and
red line) diagrams for the black hole solutions
in (a) 4 and (b) 5 dimensions with $\gamma=1/2$, $\Lambda=-1$, $k=1$.}
\label{M-rsing_L-1k1D4}
\end{figure}

In $D=4$, there are minima for the horizon radius ($r_H=3.245$) and the mass of
the solution, below which the black holes cease to exist. For the large black hole,
the spacetime approaches the GR one.
Integrating the field equations inward, we find the second derivative of the dilaton
field diverges at finite radius $r_{\rm s}$ for any black hole solution.
The formation mechanism of this fat singularity is the same as the zero cosmological
constant case. The divergent rate of the Kretschmann invariant is given by
%===========<Equation>============%
\begin{eqnarray}
\label{Kretsch_rs}
{\cal I} \sim O\biggl[\frac{1}{(r-r_{\rm s})^{4}}\biggr].
\end{eqnarray}
%=================================%
For the minimum mass solution, we find that $r_{\rm s}$ and $r_H$ coincide with each other,
and the fat singularity appears at the location of the event horizon.
We show the ${M}_0$-$r_H$ and  ${M}_0$-$r_{\rm s}$ diagrams
in Fig.~\ref{M-rsing_L-1k1D4}~(a). The global structure is the Schwarzschild-AdS type.

%--figures-------------------------------------------------------------
%--figures-------------------------------------------------------------

%===========================================
\subsubsection{$D=5$}
%===========================================

In $D=5$, there are also minima for the horizon radius ($r_H=0.805$) and
the mass of the solution.
There is the fat singularity at $r=r_{\rm s}$ for the large black holes
$r_H>2.35$ as depicted in Fig.~\ref{M-rsing_L-1k1D4}~(b). At the singularity,
the Kretschmann invariant diverges as Eq.~(\ref{Kretsch_rs}).
On the other hand, for the small black holes  $r_H<2.35$, the singularity
locates at the center. The field functions behave as
%===========<Equation>============%
\begin{eqnarray}
%\label{??}
B\sim r^{-10.9}, ~~~ e^{2\delta} \sim r^{-12.9}, ~~~ e^{\gamma\phi}\sim r^{-12.9},
\end{eqnarray}
%=================================%
and the Kretschmann invariant becomes
%===========<Equation>============%
\begin{eqnarray}
%\label{}
{\cal I} \sim O\biggl(\frac{1}{r^{25.8}}\biggr).
\end{eqnarray}
%=================================%
These divergent rates are again same as the zero cosmological constant case
[see Eqs.~\p{d5div} and \p{d5violent}].
For the minimum mass solution, the radii of the event horizon and singularity are
different.
This means that the disappearance of the black hole solution at $r_H=0.805$ is not due to
the coincidence of the internal singularity with the event horizon.
Actually, if we integrate outward from $r=0.805$ with the boundary condition
of the event horizon, the spacetime just outside of $r=0.805$ is regular.
However, continuing the integration, we find that the second derivative of
the dilaton field diverges at $r=1.119$ and the spacetime becomes singular.
Hence, the ${M}_0$-$r_H$ and  ${M}_0$-$r_{\rm s}$ curves are disconnected.

Since  the tortoise coordinate is finite
at $r=0$, the singularity is spacelike. There is no event horizon except for the black hole
horizon. The spacetime is asymptotically AdS.
Hence the global structure is again the Schwarzschild-AdS black hole type.

%===========================================
\subsubsection{$D=6$ -- $10$}
%===========================================

As in the $D=5$ case, there is the  minimum horizon radius $r_H=0.452$ for the black
hole solution in $D=6$, while we can take $r_H\to 0$ limit in $D=10$.
The outer spacetime is almost the same as that in the $k=0$ case in the large horizon
limit.
The singularity exists at the center and the spacetime is regular for $r>0$ both
in $D=6$ and $10$.

In $D=6$, the  field functions behave as Eq.~(\ref{L-1k0D6_div}) independently
of the horizon radius, and
the Kretschmann invariant diverges as
%===========<Equation>============%
\begin{eqnarray}
%\label{??}
&& {\cal I} \sim O\biggl(\frac{1}{r^{32.2}}\biggr) ~~~ (D=6).
\end{eqnarray}
%=================================%

\begin{figure}[bt]
\begin{center}
\includegraphics[width=8cm]{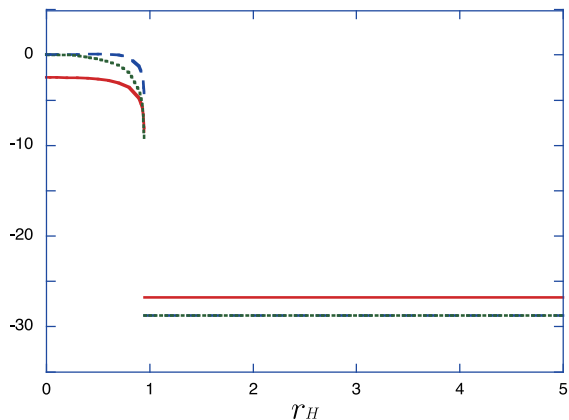}
\end{center}
\caption{The divergent rates of the field functions around the central singularity
for the black hole solution
in 10-dimensions with $\gamma=1/2$, $\Lambda=-1$, $k=1$.
We plot the exponents of $B\sim r^{a}$ (solid line), $e^{2\delta}\sim r^{b}$
(dashed line), and $e^{\gamma \phi}\sim r^{c}$ (dotted line).}
\label{rh-div_L-1k1D10}
\end{figure}

In  $D=10$, the divergent rates of the field functions depend on the horizon radius
as depicted in Fig.~\ref{rh-div_L-1k1D10}.
%--figures-------------------------------------------------------------
%--figures-------------------------------------------------------------
For $r_H>0.949$, the exponents become constant and
%===========<Equation>============%
\begin{eqnarray}
%\label{??}
&& B\sim r^{-26.8}, ~~~ e^{2\delta} \sim r^{-28.8}, ~~~ e^{\gamma\phi}
\sim r^{-28.8} ~~~ (D=10),
\end{eqnarray}
%=================================%
as in the $k=0$ case.
For $r_H<0.949$, the exponents depend on the horizon radius.
The mildest divergence of the function $B$ appears in the zero horizon limit, where
$a=-(D-5)/2,~b=0,~c=0$.
The Kretschmann invariant diverges as
%===========<Equation>============%
\begin{eqnarray}
%\label{??}
&& {\cal I} \sim O(r^{2a-4}) ~~~ (D=10),
\end{eqnarray}
%=================================%
and the mildest divergence in $D=10$ is obtained in the zero horizon limit as
%===========<Equation>============%
\begin{eqnarray}
%\label{??}
{\cal I} \sim O\biggl(\frac{1}{r^{D-1}}\biggr),
\end{eqnarray}
%=================================%
which is the same order as the non-dilatonic one.

Since  the tortoise coordinate is finite
at $r=0$, the singularity is spacelike. There is no event horizon except for the black hole
horizon. The spacetime is asymptotically AdS.
Hence the global structure of these solutions is again the  Schwarzschild-AdS black hole
type.

%%%%%%%%%%%%%%%%%%%%%%%%%%%%%%%%%%%%
\subsection{$\Lambda<0$ and $k=-1$}
\label{Lnk-1}
%%%%%%%%%%%%%%%%%%%%%%%%%%%%%%%%%%%%

In the $k=-1$ case, the qualitative properties are the same in all dimensions.
Hence we can discuss $D=4$ -- 6 and 10 cases simultaneously.

Here the basic equations have the exact solution~\cite{OT3}
%===========<Equation>============%
\begin{eqnarray}
%\label{??}
\phi\equiv \phi_0={\rm constant},~~~ \delta\equiv 0, ~~~B={b}_2r^2-1.
\end{eqnarray}
%=================================%
This solution has a black hole event horizon at $r=1/\sqrt{{b}_2}$
which is the AdS radius. The mass of the black hole is zero and the Kretschmann
invariant is finite everywhere.
%However, the center is singular because of
%the topological structure of the $(D-2)$-dimensional constant curvature space.
The global structure of this zero mass black hole is the
``regular AdS black hole" type.

\begin{figure}[bt]
\includegraphics[width=16cm]{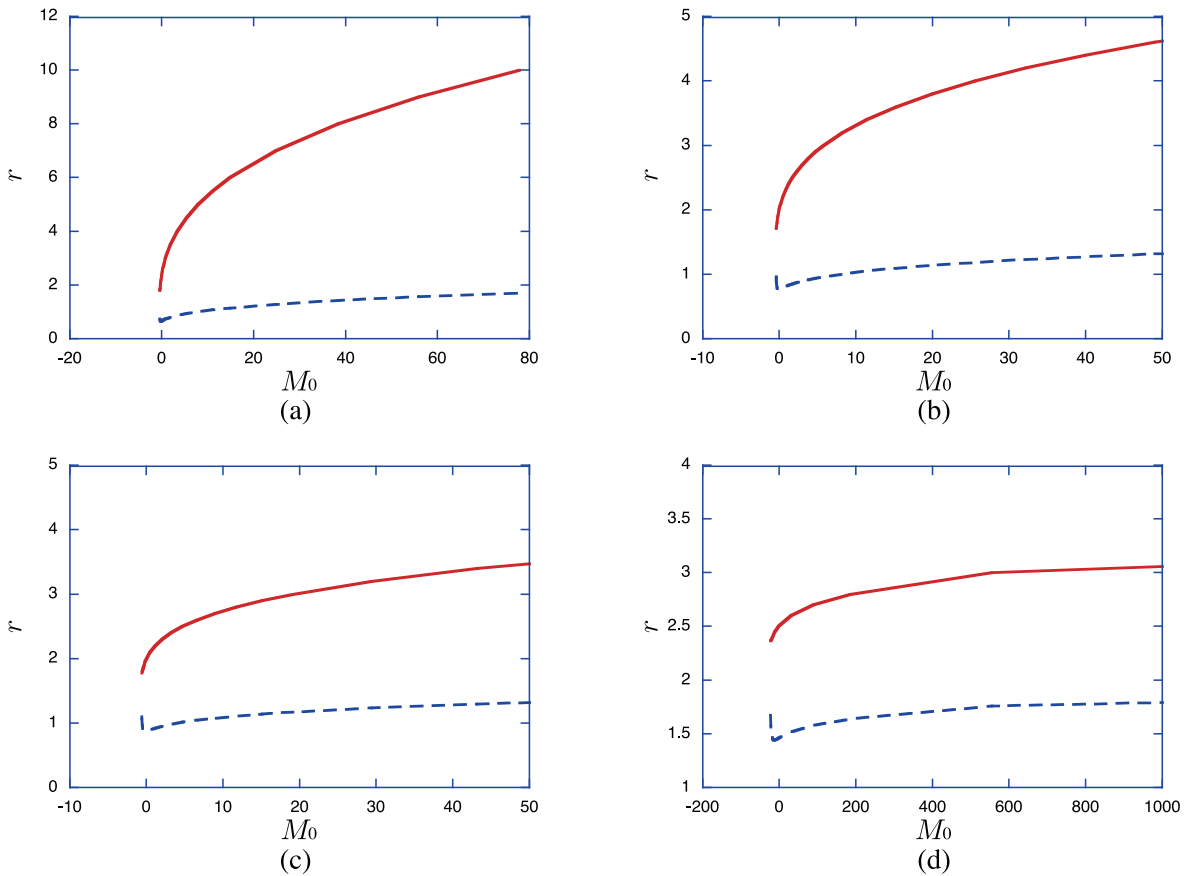}
\caption{The $r_{\rm s}$-${M}_0$ (dashed and blue line) and $r_H$-${M}_0$ (solid and
red line) diagrams for the black hole solutions
in (a) 4, (b) 5,  (c) 6, and (d) 10 dimensions with $\gamma=1/2$, $\Lambda=-1$, $k=-1$.}
\label{M-rsing_L-1k-1D}
\end{figure}

Except for the zero mass black hole solution, the singularity appears
at the nonzero finite radius $r_s$ where the second derivative of the dilaton
field $\phi''$ diverges. Fig.~\ref{M-rsing_L-1k-1D} shows the radii $r_s$ and $r_H$
as functions of the black hole mass. The behavior
of the field functions is given by Eq.~(\ref{fields_r_s}), and the Kretschmann
invariant becomes
%===========<Equation>============%
\begin{eqnarray}
%\label{??}
{\cal I} \sim O\biggl[\frac{1}{(r-r_s)^{4}}\biggr].
\end{eqnarray}
%=================================%
In the non-dilatonic case, the Kretschmann invariant is ${\cal I} \sim O(r^{-(D-1)})$
for the central singularity of the positive mass black hole while
${\cal I} \sim O\bigl[(r-r_b)^{-3}\bigr]$  for the branch singularity of the negative
mass black hole.
Compared with the non-dilatonic case, the diverging behavior becomes mild for the
positive mass solutions in $D=6$ -- 10. On the other hand, it is stronger for the positive
mass solutions in $D=4$ and negative mass solutions in all dimensions.
%--figures-------------------------------------------------------------
%--figures-------------------------------------------------------------

The black hole solution has critical horizon radius $r_H=1.807,~1.712,~1.777$, and 2.364,
for $D=4,~5,~6$, and 10, respectively, below which there is no black hole solution.
At the critical horizon radius, the radii of the horizon and the singularity coincide
and the horizon becomes singular. It appears in Fig.~\ref{M-rsing_L-1k-1D} as if
the curves of $r_H$ and $r_{\rm s}$ is disconnected. However, this is just due to
the difficulty of the numerical analysis around the critical horizon radius,
and they should be connected.

Since  the tortoise coordinate is finite
at $r=0$, the singularity is spacelike. There is no event horizon except for the black hole
horizon. The spacetime is asymptotically AdS.
Hence the global structure of these solutions is again the  Schwarzschild-AdS black hole type.

%%%%%%%%%%%%%%%%%%%%%%%%%%%%%%%%%%%%%%%
\section{Conclusions and Discussions}
\label{CD}
%%%%%%%%%%%%%%%%%%%%%%%%%%%%%%%%%%%%%%%

We have investigated how the dilaton field affects the global structure of the black
hole solution in Einstein-Gauss-Bonnet gravity. In this paper the system includes
the zero and negative cosmological constant and the spacetime is assumed to have
spherical, planer, and hyperbolic symmetry.
%The global structure of the spacetime is
%determined by the asymptotic structure, event horizons and the singularity.
The global structure of the solution is changed by the effect of the dilaton field
drastically.
In the non-dilatonic system, the solutions have six different types of
global structure while there are just three types in the dilatonic system.
They are the Schwarzschild type for the zero cosmological constant,
the Schwarzschild-AdS type in the negative cosmological constant case,
and the ``regular AdS black hole" type for the zero mass black hole.

One of the important issues is whether the dilaton field makes the strength of the
singularity mild or not.
We have investigated the Kretschmann invariant around the singularity in detail.
We have found that the singularity is classified into three types.
One is the central singularity around which the divergent rate of the Kretschmann
invariant does not depend on the horizon radius as in the non-dilatonic case.
The second one is the central singularity around which the divergent rate depends
on the horizon radius. The third one is the fat singularity, which locates at
the non-zero finite radius. All types of singularities are spacelike.
The divergent rates of the  Kretschmann invariant are summarized in
Table~\ref{table_sammary_0} and \ref{table_sammary_-1}.
The first type of the central singularity has violently divergent behavior with the rate
${\cal I}\sim r^{-6.36D+5.94}$.
On the other hand, the fat singularity has rather mild behavior.
We find that there are cases where the singularity becomes mild
by the presence of the dilaton field in the system.
There is also some dependence on the dilaton coupling whether this mild fat singularity
appears or not. We find an indication that it is more likely to appear for the
larger dilaton coupling.

As for the event horizon, the black holes solution has just one horizon in the dilatonic
system generically. It is due to the structure of the equation of the dilaton field.
The equation of the dilaton field (\ref{fe3}) is rewritten as
%--------
\begin{eqnarray}
B\phi^{\prime \prime }
=
-\frac{1}{r^{D-2}e^{-\delta}}\bigl(r^{D-2}e^{-\delta}B \bigr)' \phi'
+\bigl(\gamma e^{-\gamma\phi}  R_{GB}^2
+\lambda {\Lambda}e^{\lambda\phi} \bigr).
\label{horizoneq}
\end{eqnarray}
%--------
At the horizon $B=0$, this equation, when solved for $\phi''$, is singular and
the right hand side should vanish. By this condition, $\phi$ and $\phi'$ cannot
be free but should be related with each other at the horizon. As a result,
there is one free parameter $\phi(r_H)$ at the black hole event horizon.
This parameter is determined by the asymptotic condition at $r \to \infty$.
However, if there were other event horizons such as the inner horizon and
the cosmological horizon, we should have additional conditions between $\phi$
and $\phi'$ at these horizons.
It is difficult, if not impossible, to satisfy the conditions at some horizons and
at infinity simultaneously in a generic occasions. For this reason,
our solutions have only single horizon i.e., black hole event horizon generically.
This situation seems to remain the same even if we add the charge to the black hole.

%--table------------------------------------------------------------------------------
\begin{table}
\begin{center}
\caption{
Summary of the divergent rate of the Kretschmann invariant ${\cal I}$ around
the singularity in the $\Lambda =0$ case. We also show those in GR and non-dilatonic
cases for comparison.
}
\label{table_sammary_0}
\begin{tabular}{c|c||cc|c|c}
\hline
$k$ & $D$  & \multicolumn{2}{c|}{$\cal{I}$} & ~$\cal{I}$ (GR)~
 & ~$\cal{I}$ (non-dilatonic)~ \\
\hline %------------------------------------------
&  4 &$(r-r_{\rm s})^{-4}$    &    &    &       \\
\cline{2-4} %------------------------------------------
 & $ $  & $r^{-25.8}$   & $(M_0 < 7.46)$  &      &   \\
                       &  \raisebox{1.5ex}[0pt]{$5 $}
                       & ~~$(r-r_{\rm s})^{-4}$  &  $(M_0 > 7.46)$  &   &       \\
\cline{2-4} %------------------------------------------
\raisebox{1.5ex}[0pt]{$1$} &   & $r^{-32.2}$   & $(5.18<M_0 < 120.9)$~~
  & \raisebox{1.5ex}[0pt]{$r^{-(2D-2)}$}     & \raisebox{1.5ex}[0pt]{$r^{-(D-1)}$}   \\
                       &  \raisebox{1.5ex}[0pt]{$6 $}
                       & $r^{-(D-1)}$  &  (otherwise)  &   &       \\
\cline{2-4} %------------------------------------------
                      & 10 & ~~$r^{-(D-1)}<{\cal I}<r^{-11.9}$  &  &  &    \\
\hline %------------------------------------------
\end{tabular}
\end{center}
\end{table}

%--table------------------------------------------------------------------------------

%--table------------------------------------------------------------------------------
\begin{table}
\begin{center}
\caption{
Summary of the divergent rate of the Kretschmann invariant ${\cal I}$ around
the singularity in the $\Lambda =-1$ case.
We also show those in GR and non-dilatonic cases for comparison.
}
\label{table_sammary_-1}
\begin{tabular}{c|c||cc|cc|cc}
\hline
$k$ & $D$  &  \multicolumn{2}{c|}{\mbox{\hspace{26mm}}$\cal{I}$
\mbox{\hspace{26mm}}} & \multicolumn{2}{c|}{ $\cal{I}$ (GR) } &
\multicolumn{2}{c}{$\cal{I}$ (non-dilatonic)}  \\
\hline %-------------------------------------------------------------------------------
&  4 &$(r-r_{\rm s})^{-4}$    &    &    &    &   \multicolumn{2}{c}{$r^{-6}$}    \\
\cline{2-4}\cline{7-8} %------------------------------------------
 & $ $  & $r^{-25.8}$   & $(M_0 < 8.39)$  &     &     &  & \\
                       &  \raisebox{1.5ex}[0pt]{$5 $}
                       & $(r-r_{\rm s})^{-4}$  &  $(M_0 > 8.39)$  &   &  &      &   \\
\cline{2-4} %------------------------------------------
\raisebox{1.5ex}[0pt]{$1$}  & 6 & $r^{-32.2}$  &  &&      &
\multicolumn{2}{c}{$r^{-(D-1)}$}   \\
\cline{2-4} %------------------------------------------
 &   & $r^{-(D-1)}<{\cal I}<r^{-21.0}$   & $(M_0 < 1.93)$  &  & &  &  \\
&  \raisebox{1.5ex}[0pt]{$10$}  & $r^{-57.6}$  &  $(M_0 > 1.93)$  &
\multicolumn{2}{c|}{\raisebox{1.5ex}[0pt]{$r^{-(2D-2)}$}}    &   \\
\cline{1-4}\cline{7-8} %----------------------------------------------------------------
 & 4  &   &   &     &   &  \multicolumn{2}{c}{$r^{-6}$} \\
\cline{2-2}\cline{7-8}
 & 5  &  \raisebox{1.5ex}[0pt]{$(r-r_{\rm s})^{-3}$}   &   &     &    &    &\\
\cline{2-4}
 \raisebox{1.5ex}[0pt]{0}  & 6  & $r^{-32.2}$  &   &     &  &
 \multicolumn{2}{c}{$r^{-(D-1)}$} \\
\cline{2-4}
 & 10  &  $r^{-57.6}$ &   &     &  &    & \\
\hline %--------------------------------------------------------------------------------
 & 4  &   &   &     & &  \multicolumn{2}{c}{$r^{-6}$  $(M_0 \ne 0)$,  $r^{0}$
 $(M_0 = 0)$} \\
\cline{2-2}\cline{7-8}
 & 5  &  $(r-r_{\rm s})^{-4}$   & $(M_0 \ne 0)$  &  $r^{-(2D-2)}$ & $(M_0 \ne 0)$ &
$(r-r_{\rm b})^{-3}$ & $(M_0 < 0)$  \\
\cline{2-2}
 & 6  & $r^{0}$  & $(M_0 = 0)$  &   $r^{0}$  &  $(M_0 = 0)$  &  $r^{0}$  &  $(M_0 = 0)$  \\
\cline{2-2}
 & 10  &   &   &      &  &  $r^{-(D-1)}$ &  $(M_0 > 0)$  \\
\hline
\end{tabular}
\end{center}
\end{table}
%--table------------------------------------------------------------

\section*{Acknowledgements}
We would like to thank Tamiaki Yoneya for useful discussions which motivated
the present work, and Hideki Maeda for valuable discussions.
This work was supported in part by the Grant-in-Aid for
Scientific Research Fund of the JSPS Nos. 20540283, 21$\cdot$09225, 21740195, 22244030 and 22540293.

%%%%%%%%%%%%%%%%%%%%%%%%%%%%%%%%%%%%%%%%%%%%%%%%%%%%%%%%%%%%%%%

\end{document}